\pdfoutput=1

\documentclass[conference]{IEEEtran}

\usepackage{cite}
\usepackage{amsmath,amssymb,amsfonts}
\usepackage{algorithmic}
\usepackage{graphicx}
\usepackage{textcomp}
\usepackage{xcolor}
\def\BibTeX{{\rm B\kern-.05em{\sc i\kern-.025em b}\kern-.08em
    T\kern-.1667em\lower.7ex\hbox{E}\kern-.125emX}}
\begin{document}

\pagenumbering{roman}
\pagestyle{plain}
\title{Stochastic Estimated Risk for Storage Capacity\\
}

\author{\IEEEauthorblockN{Revathi Anil Kumar}
\IEEEauthorblockA{\textit{Data Scientist} \\
\textit{Nutanix}\\
San Jose, California \\
revathi.anilkumar@nutanix.com}
\and
\IEEEauthorblockN{Mark Chamness}
\IEEEauthorblockA{\textit{Director, Data Science} \\
\textit{Nutanix}\\
San Jose, California \\
mark.chamness@nutanix.com}
}

\maketitle

\begin{abstract}
Managing data storage growth is of crucial importance to businesses. Poor practices can lead to large data and financial losses. Access to storage information along with timely action, or capacity forecasting, are essential to avoid these losses. In addition, ensuring high accuracy of capacity forecast estimates along with ease of interpretability plays an important role for any customer facing tool. 
In this paper, we introduce Stochastic Estimated Risk (SER), a tool developed at Nutanix that has been in production. SER shifts the focus from forecasting a single estimate for date of attaining full capacity to predicting the risk associated with running out of storage capacity. Using a Brownian motion with drift model, SER estimates the probability that a system will run out of capacity within a specific time frame. Our results showed that a probabilistic approach is more accurate and credible, for systems with non-linear patterns, compared to a regression or ensemble forecasting models. 
\end{abstract}


\section{Introduction}
As data utilization continues to increase, it is crucial for businesses to adopt efficient storage practices. A reactive approach, where system administrators take action after a system has hit full capacity can prove to be a costly exercise, both financially and operationally. With the advent of machine learning and predictive modeling, we can employ such methods to enable a more proactive approach to managing storage capacity. 

Applying predictive modeling to storage capacity has been done by methods such as Chamness Regression\cite{Chamness} and Symantec’s Soothsayer \cite{Soothsayer}. Both approaches have contributed significantly to this field, and relied on datasets that often exhibited linear capacity utilization trajectories. At Nutanix we observed a combination of trends including, linear, non-linear, continuous and discontinuous storage growth patterns. Its unclear what causes these changes in workload behaviors. Potentially the underlying causes can be a combination of system and human behavioral changes. Initially, we relied on a simple linear growth assumption to build a model that predicts when a system would run out of capacity. This estimated forecast was utilized by customer support and customers to drive decisions related to managing storage capacity.The model was inadequate for non-linear workloads and this provided motivation to develop a completely new approach. 

Our goal was to conduct an experiment to identify the error across a set of forecasting models. We initiated our experiment with the hypothesis that error across three models would be the same. The three models compared were a piece-wise linear regression, an ensemble model and a naive model. The naive model assumes that capacity remains constant from the last observed value. For a model to be considered, it must be as good as the performance of the naive model. However, analysis of total error showed  that the naive model performed better than both models One probable reason for this could be the non deterministic behavior associated with storage utilization. A combination of system and behavioral changes (i.e. deleting data) can result in systems without predictable patterns. 

The goal of existing models is to forecast the time to full capacity. The uncertainty associated with the system workloads often result in poor estimates. The aim of providing a forecast is to help system administrators avoid losses associated with reaching full capacity. Reframing this problem in terms of statistical risk analysis, the outcome can be redefined as a probability. Statistical risk assessment models are often used to estimate the probability of an event. By providing a probability of failure, instead of a point forecast, we can help system administrators mitigate the risk associated with full capacity.  This can also help accommodate for risk aversion preferences. For example, development systems might tolerate a 50\% probability of failure, while production systems might only tolerate a 10\% probability of failure. With the randomness of storage capacity and the aim of mitigating risk, we considered how other industries addressed similar issues. Within the financial services industry, geometric Brownian motion was used to develop the Black-Scholes model to price put and call options. 

Adopting a similar approach, we introduce Stochastic Estimated Risk (SER), a probabilistic model that moves from estimating the date of running out of capacity to estimating the probability of failure. For a specified time interval, SER estimates the probability that a system could run out of capacity. SER uses stochastic processes based on Brownian motion with drift to provide probability estimates,

An outline of this paper is as follows. Recent work done in this field has been summarized in Section II. Section III compares two forecasting methods against a naive model: piece-wise regression and an ensemble of forecasting models. Section IV outlines the new model SER, and its application to risk prediction. Section V discusses a visual comparison between SER and a linear regression model, Section VI discusses the accuracy estimates for the Brownian motion model and Section VII reviews optimizing business decisions using SER.

\section{Related Work}
Multivariate Adaptive Regression Splines (MARSplines) popularized by Friedman\cite{Friedman}, partitions an input space into regions and builds a best fit regression model based on the region. Chamness regression, a method developed at EMC, to forecast workloads, is an extension of MARSplines. This model predicts the date that a system is expected to run out of capacity\cite{Chamness}. The Chamness paper utilizes a piece-wise linear regression method to select the best subset of historical data for prediction. A point estimate, generated by assuming a linear trend in the system, is inaccurate for systems experiencing non-linear trends. The uncertainty associated with storage systems often requires the use of models that avoid assuming or imposing any particular relationship between the predictor and outcome of interest. 

Symantec developed Soothsayer, a model that uses an ensemble model and provides a confidence interval for reaching full capacity\cite{Soothsayer}.  Instead of employing observations of data capacity, a model is built using the size of backup and deduplication ratios. The paper outlines three models - ARIMA, Stochastic Model and a merged model combining ARIMA and Stochastic techniques. The paper recognized that systems exhibit distinct capacity utilization workloads. Systems are broadly classified into three groups: linear, trending and stratified. Forecasting models are applied to these three groups. However, regardless of the group, underlying growth is assumed to be linear across all groups. Since these models do not account for non-linear workloads, they will result in poor estimates. 

At Nutanix, we noticed trends in capacity were highly unpredictable. This may be potentially due to customer activities such as deleting data, migrating VM’s or a change in utilization patterns. In this paper, we first quantify the efficacy of existing linear methods by comparing models used in the industry to a naive model. These results helped validate the need to reframe the question. One of the aims of system administrators is to reduce risk associated with running out of capacity, and by using a method that provides a probabilistic estimate, the risk associated with storage utilization can be estimated.  

Brownian motion is one of the methods used when predicting with uncertain elements. Brownian motion, first observed by Robert Brown, was used to describe the random motion of particles as a result of collisions. There appeared a common trend where Brownian motion methods are utilized when there is a high degree of randomness associated with the event or object under study. Some applications include but are not limited to, utilizing Brownian motion to classify medical images with a high degree of complexity\cite{Oczeretko}, a probability model based on Brownian motion which modeled rupture times at the source of recurrent earthquakes\cite{Matthews}. The applications in the financial industry are more closely related with the methods carried out in this paper. Osborne’s paper on Brownian motion in the stock market, establishes a correspondence between the stock market and the movement of particles, and states that statistical methods applied to study the latter can be used for the former\cite{Osborne}. More popularly, the Black Scholes Model used to price put and call options assumes a geometric Brownian motion\cite{Black}. 

\begin{figure}[htbp]	
	\includegraphics[width=80mm,scale=0.8]{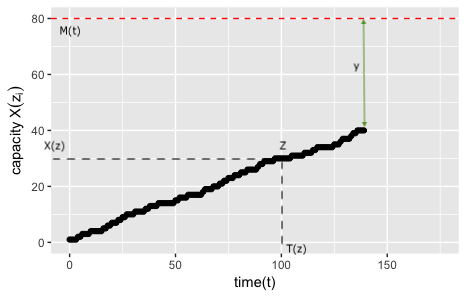}
	\caption{Modeling capacity utilization as a Stochastic process where red dotted line refers to 100\% capacity}
	\label{fig}
\end{figure}

Assuming that the growth in storage follows a continuous Markov process. The transitions in capacity consists of various states such as adding more storage, deleting data, moving data and  reaching full capacity. Modeled as a Markov process, and identifying the position and variance associated with storage within a certain time period, we use a Brownian motion with drift model to estimate the likelihood that a system would attain full capacity within a certain time period. This process is similar to various hitting time problems, where the objective is to estimate if an object attains a maximum value within a time frame. Mathematically, let $X(z)$ be a Brownian motion process for denoting storage utilization for all $z \geq 0$. $X(z)$ would be the storage growth observed at time $T(z)$. Given this Brownian process, the maximal value of the process up to time $t$, with a drift coefficient of $\mu$ and variance $\sigma^2$, can be defined as:

\begin{equation}
	M(t) = max_{0 \leq p \leq t} X(p) 
\end{equation}

To estimate the risk associated with running out of capacity, or hitting the maximum capacity within a given time frame, we model risk mathematically as the probability that maximal value of capacity at time $t$, $M(t)$ is greater than or equal to $y$, which is the difference between total capacity and capacity used. This is demonstrated in Figure 1 where $y$ is the difference between most recent capacity used and total capacity. Ross\cite{Ross} defines the probability using the following formula:

\begin{equation}
	P(M(t) \geq y) = e^{2y\mu/\sigma^2} \overline{\Phi}(\dfrac{y + \mu t}{\sigma\sqrt{t}}) +  \overline{\Phi}(\dfrac{y - \mu t}{\sigma\sqrt{t}})
\end{equation}

We reframed the question and utilize a method that makes no assumption of linear trends. In addition, this method estimates the likelihood of hitting maximum capacity, within a given time interval regardless of whether maximum capacity was hit at the beginning or towards the end. 

\vspace{5mm}

\section{Experiment Section}

\subsection{Model Description}

For this section we outline the experimental method and discuss results for an experiment. This experiment compares methods that have been used to forecast when a cluster is estimated to run out of capacity, to a naive model. Before we delve into the methods, we describe the three models and the assumptions/conditions for each model. 
\vspace{5mm}

\subsubsection{Piecewise Linear Regression Model}

We emulated the Chamness regression\cite{Chamness} model based on piece-wise linear regression. To reduce the error rate associated with fitting a linear regression model, the best subset is selected, which is most representative of recent changes in storage utilization, and a model is fit to this subset. To determine the best fit subset, the model uses $R^2$, defined as the regression sum of squares divided by the total sum of squares. Following are additional conditions introduced into the model based on the paper\cite{Chamness} :
\begin{itemize}
	\item Threshold for $R^2$ - A threshold is set for $R^2$ of 70\%, a $R^2$ value below this suggests a poor fit to the datasets
	\item Positive slope - The model checks for a positive correlation. A model with zero or negative slope cannot be used to forecast utilization
	\item Sufficient observations - Based on the Chamness paper\cite{Chamness}  15 days is identified as sufficient information to identify patterns or forecast storage	
\end{itemize}

\vspace{5mm}
\subsubsection{Ensemble of models}
Next we looked at an ensemble of models that uses a tournament to select the best model for each storage workload and historical time range.	An ensemble of four models -  Seasonal decomposition of time series by Loess (STL), Linear regression, Theta forecast and ARIMA. Followed by a cross validation procedure, the error is computed for each of the methods and top two models were selected. Based on the results, forecasts of the top two models were combined based on weights computed by the ratio of errors. 	

\vspace{5mm}
\subsubsection{Naive Model}
The naive model takes the amount of storage on the last day and forecasts that the capacity for next 7 days is the same. This is considered to be a stochastic process with a martingale property. A stochastic process, for storage values ${X_t: t =0,1,...}$ for various points of time $t$, is considered a martingale if, for $t =0,1,...$,
\vspace{5mm}
\begin{itemize}
	\item $E[|X_t|] < \infty$
	\item $E[X_{t+1}| X_0,...,X_t] = X_n$
\end{itemize}
\vspace{5mm}

\subsection{Method Description}
The dataset consists of observations of storage utilization from Nutanix customers. When customers enable Pulse, we are able to monitor and forecast their storage utilization. We applied the models to those systems that were actively utilized. For such systems, we gathered sufficient utilization observations to measure and compare accuracy. We limited our analysis to systems that are over 10\% capacity utilization and have a variance of greater than 0.01\% in the last 10 days. We carried out a restricted experiment using a random sample of data consisting of 1000 systems. 

To identify if either the piece-wise linear regression or ensemble model is better than the naive model, we designed an experiment to test for difference in means for each method against the naive method. Assuming the naive method is the simplest one to forecast capacity, we identified if either the piece-wise linear regression or ensemble method is better than the naive. To test this, we ran three models on the sample data. For each model, we performed cross validation by forecasting capacity on the test set for 7 most recent days of data. We computed the mean absolute percentage error (MAPE) on the test set. Based on the conditions mentioned in the model description section, each model might not compute an error for every cluster. We ran hypothesis testing on those systems that have produced forecasts across all three models, which is 70\% of the dataset. We set up 2 Welch's t-test such that for any given method $m, m\in (1,2)$, i.e. 1 denoting piece-wise linear regression and 2 a forecasting ensemble, we took the average MAPE and defined the null hypothesis as: 

\vspace{5mm}
$H_0: \mu_{naiv\acute{e}} - \mu_{m} = 0$

\vspace{5mm}
And the alternative hypothesis as: 

$H_a: \mu_{naiv\acute{e}} - \mu_{m} \not= 0$

\begin{table}[htbp]

	\begin{center}
		\begin{tabular}{|c|c|c|c|c|}
			\hline
			\cline{2-5} 
			\textbf{Method} & \textbf{\textit{$\mu_{naiv\acute{e}}$}}& \textbf{\textit{$\mu_{m}$}}& \textbf{\textit{p-value}} &\textbf{\textit{95\% CI}} \\
			\hline
			Piecewise Linear Regression& 0.028 & 0.035& 0.03 & -0.034, 0.018  \\
			\hline
			Ensemble Model & 0.028 & 0.029 & 0.49 & -0.031, 0.022 \\
			\hline
		\end{tabular}
		\label{tab1}
		\caption{Results for Hypothesis testing}
	\end{center}
\end{table}

Table 1 depicts results obtained from the analysis. For piece-wise linear regression, the naive model had an average MAPE of 2\% compared to a 3\% average error for the piece-wise linear regression model, suggesting that the naive model was more accurate at forecasting when a cluster is expected to run out of capacity, with a statistically significant difference in means. The error rate for the naive model was slightly lower than the error rate for the ensemble model, however the difference in means was not statistically significant. 

We ran 100 iterations of the same experiment, drawing 1000 random samples each time.  We observed a large variance in average error computed across different experiments. While the naive model performed better on average, the large variance further strengthens our concern on the uncertain nature of storage trends and raises the question whether commonly used forecasting models can be applied to storage systems. 

\vspace{5mm}
\section{Stochastic Estimated Risk using Brownian motion with drift}

\subsection{Assumptions of a Brownian motion with drift}

In its simplest form, Brownian motion is considered an infinitesimal random walk. 

We implemented a Brownian Motion with drift model to estimate the risk associated with running out of capacity within a certain time period. As described by Karlin and Taylor\cite{Karlin} Brownian motion with drift is considered a stochastic process ${X_t: t >0}$ with the following properties:
\begin{itemize}
	\item Every increment $X(t +s) - X(s)$ is normally distributed with a mean $\mu t$ and variance $\sigma^2t$
	\item For every pair of disjoint time intervals $[t_1, t_2], [t_3,t_4]$, say $t_1 <  t_2 \leq t_3 < t_4$, the increments $X(t_4) - X(t_3)$ and $X(t_2) - X(t_1)$ are independent random variables with normal distributions and mean $\mu t$ variance $\sigma^2t$, and similarly for a disjoint time intervals where n is an arbitrary positive integer
	\item $X(0) = 0 $ and $X(t)$ is continuous at $t = 0$
\end{itemize}

\vspace{5mm}

\subsection{Extending Brownian motion to Capacity Utilization}
Modeling storage growth as a Brownian process with drift, we can estimate the probability of attaining full capacity within a given time period. The maximal value of the process within a time interval $t$ is defined in Equation 1 as:

\vspace{5mm}
$ M(t) = max_{0 \leq u \leq t} X(u) $

\vspace{5mm}

To determine the probability that maximal of a Brownian motion with drift model within a given time interval, would be greater than $y$, we use the formula provided by Ross\cite{Ross} in the Introduction to Probability Models. For any given Brownian motion process, we can compute the likelihood using the following equation:

\begin{equation}
P(M(t) \geq y) = e^{2y\mu/\sigma^2} \overline{\Phi}(\dfrac{y + \mu t}{\sigma\sqrt{t}}) +  \overline{\Phi}(\dfrac{y - \mu t}{\sigma\sqrt{t}})
\end{equation}

Where the inverse cumulative distribution function is: 

\begin{equation}
P(Z > x) = \overline{\Phi}(x) = 1 - \Phi(x)
\end{equation}

Applying this to storage utilization, we are interested in estimating the likelihood that a system would run out of capacity within a time period $t$, say 30 days in our example. The maximum value $M(t)$ refers to the total capacity attained by the system within time period $t$. For a time period $0 \leq z \leq t$, $z$ refers to the most recent data point available on customer storage patterns. We are interested in utilizing time series data up to point $z$, to determine the likelihood of full capacity within a time period $t$. $X(z)$ is characterized as the capacity at time $z$, which follows a Brownian motion process. 

The rate at which the difference between $X(z)$ and $M(t)$ reduces would determine the likelihood that a system would run out of capacity within $t$ time.  Let $y = M(t) - X(z)$, then the probability that total capacity of the system $M(x_t)$ is greater than or equal to the distance traveled $y$ is equal to Equation (3). 

As stated by Karlin and Taylor\cite{Karlin} every increment of capacity or $X(z)$ is considered to follow a normal distribution with mean $\mu t$ and variance $\sigma^2 t$. Values for the risk estimate are obtained from a normal distribution $\Phi$ as denoted in Equation 4. 

\subsection{Parameter estimation - Drift and Variance}

To estimate the drift and variance, $\mu$ and $\sigma^2$, we utilize the properties of Brownian motion with drift model which states that the increment $X(t +s) - X(s)$is normally distributed with a mean $\mu t$ and variance $\sigma^2 t$. For every storage system we determine the incremental changes i.e $X(z) - X(z-1)$. Based on the properties stated for a system with vector $z_i, i \in [0,...,n]$ containing storage, drift and variance can be computed using the difference $D_i$ defined as:

\begin{equation}
D_i = X(z_i) - X(z_{i-t}) 
\end{equation}
\begin{equation}
\mu = \dfrac{1}{t}\sum D_i
\end{equation}
\begin{equation}
\sigma^2 = \frac{1}{t}\dfrac{\sum (D_i - \overline{D})^2}{n-1}
\end{equation}

Following which we test for normality using the Welch’s t-test. With a significant p-value estimate of less than 0.05, we ascertain the properties of a Brownian motion with drift model. If the data met the properties specified, we proceeded to compute the drift and variance using the equations mentioned in 6 and 7. 

\vspace{5mm}
\section{Visual Comparison}

Nutanix systems often exhibit unique storage trends and a linear regression or ensemble model provided poor forecast estimates for such systems. In this section we provide examples of such systems and show how the Brownian motion model adapts to such changes while still providing reliable risk estimates for storage utilization.

We first look at systems that have very low variance on capacity utilization(less than 0.05). Variance of the system is computed utilizing the method mentioned in previous section. Figure 2 shows a system with very low variance across 120+ days. We utilize a piece-wise linear regression model, which selects the best subset and forecasts when the system is expected to reach a 100\% capacity (dashed red line). Based on system trends, a piece-wise linear regression model uses a best fit line with a gradual negative slope. In such a case it appears that the system will not run out of capacity at any point in the future. Figure 3 shows the Brownian motion with drift model applied to the same system. Each vertical line depicts the risk associated with running out of capacity in the next 30, 60 and 90 days. While the probability of failure is very close to zero at 30 days, this gradually increases. Considering that the drift and variance are parameters, the model adapts trends in storage patterns and incorporates the drift and variance. 

\begin{figure}[htbp]	
	\includegraphics[width=80mm,scale=0.8]{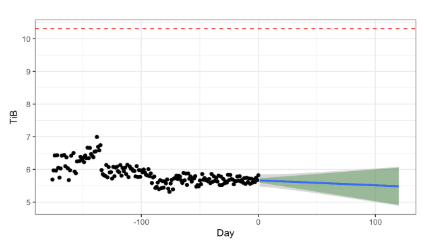}
	\caption{Piece-wise Linear Regression on low variance cluster where red dotted line refers to 100\% capacity}
	\label{fig}
\end{figure}

\begin{figure}[htbp]
	\includegraphics[width=80mm,scale=0.8]{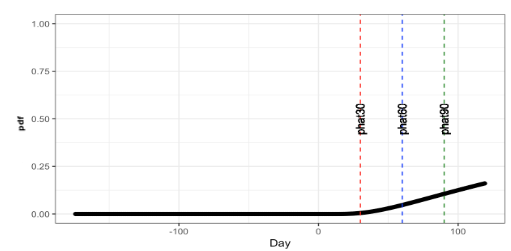}
	\caption{Brownian motion model on low variance cluster where red dotted line refers to 100\% capacity}
	\label{fig}
\end{figure}

Next we considered a system with higher variance. Figures 4 and 5 show a system that is classified as high utilization system, since its capacity utilization within the last 100 days has been above 50\% of capacity. Figure 4 shows a piece-wise linear regression model with the best fit line extended to full capacity. The system is expected to reach a 100\% within the next 50 days. We applied the Brownian motion with drift model to the same data as seen in Figure 5. The results showed that the system was at 88\% risk of running out of capacity within the next 30 days. 

\begin{figure}[htbp]
	\includegraphics[width=80mm,scale=0.8]{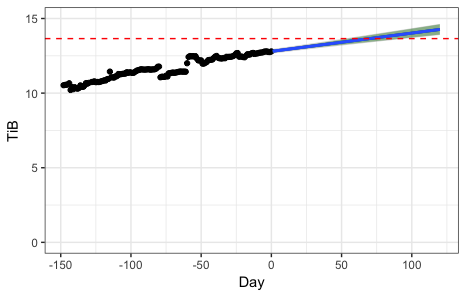}
	\caption{Piece-wise Linear Regression on high variance cluster where red dotted line refers to 100\% capacity}
	\label{fig}
\end{figure}

\begin{figure}[htbp]
	\includegraphics[width=80mm,scale=0.8]{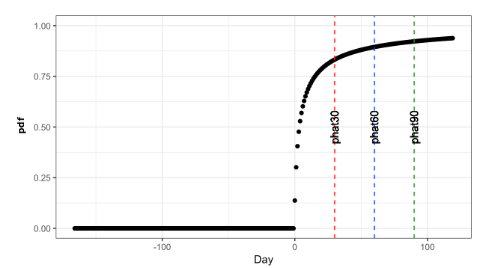}
	\caption{Brownian motion model on high variance cluster where red dotted line refers to 100\% capacity}
	\label{fig}
\end{figure}

\vspace{5mm}

\section{Measuring Accuracy}

We measured the performance of Brownian motion model against existing models such as the piece-wise regression model. The aim of each model is to optimize the decision rule which is to accurately identify whether a customer is at risk of running out of capacity. The business outcome is to decide whether or not to call a customer that is at risk of running out of capacity. We translated our model outcomes into a Bernoulli outcome such that, for every customer there is a quantity $d(x) \in [0,1]$ denoting risk where $ d(x) =1$ suggests a customer is at risk, and $d(x) = 0$ suggests a customer is not at risk.

The dataset consists of a restricted sample of systems that have at least 60 days of data, for training and testing purposes. As well as systems that are between 50\% to 70\% capacity utilization. We noticed that systems above 50\% capacity utilization are most susceptible to changes in capacity over a 30 day period, which is required to capture the accuracy of the model. 

For the purpose of measuring accuracy, we looked for the ability of the model to detect small changes in capacity. We wanted to ensure that we have enough systems that have seen changes in storage over a +60 day period. Based on this as well as increasing the size of the dataset, we looked at a 2\% increase in capacity over a 30 day period. Hence, systems that had a 2\% increase in capacity were classified as at risk, and systems that did not have a 2\% increase in capacity were classified as not at risk. 

We split the data into training and testing such that the most recent 30 days were used for testing and the remaining were used to train the model. The following set of rules were followed for each model:

\subsection{Brownian motion with drift}
\begin{itemize}
	\item  The data included a vector of storage capacity across 60+ days
	\item With the training set, we computed the probability that a system would increase by 2\% in the next 30 days using Equation 3
	\item For methods used to estimate risk, it is common practice to set a threshold $t$ for the probability computed, $p(x)$. When $p(x) > t$,  the decision on $d(x) = 1$
	\item In our case, systems that are expected to be at risk i.e. increase by 2\% in the next 30 days, are identified with $d(x) = 1$. The optimal threshold $t$ was determined using the true positive and false positive rate. At $t = 60\%$, we retain a higher true positive rate in comparison to the true negative rate. 
	\item From the test set, we look for at least 2\% increase in capacity across the 30 day period. If the system has experienced at least 2\% increase in capacity, $d(x) = 1$ for that system. 	
\end{itemize}

\subsection{Piecewise linear regression}
\begin{itemize}
	\item The data table included two columns, storage capacity and day(the difference between last date of the training set and each date)
	\item The piece-wise linear regression model was run on the training set with storage capacity as the predictor. Based on the coefficients, we computed the number of days until a 2\% increase in capacity. Let the capacity on last date of the training set equal $C_n$, then the formula used to compute remaining days ($D_r$) is as follows:
	
	\vspace{5mm}
	$D_r = \left(\frac{2\%\ increase\ from\ C_n}{\beta_{day}} \right)$
	
	\vspace{5mm}
	\item For systems that are expected to see a 2\% increase in capacity in 30 days, the remaining days should be equal to or less than 30 i.e. if $D_r \le 30$,  $d(x) =1$
	\item Finally, similar to the Brownian motion model, we identify if a system has actually seen a 2\% increase in capacity over a 30 day period. 
\end{itemize}
	
\vspace{5mm}

By carrying out the steps mentioned in A and B, we obtained a pair of vectors for each model - actual versus test results. With these vectors, we computed the true positives, false positives, true negatives, false negatives, precision and accuracy estimates. 

With a sample of 700+ systems, we ran the above analysis on all clusters and obtained the following results. 
\begin{itemize}
	\item The Brownian motion model provided estimates for all 700+ systems, where as the piece-wise linear regression model provided estimates for 526 systems
	\item Table 2 provides a confusion matrix from 700+ clusters for the Brownian motion model. Table 3 and Table 4 look into 525 systems that are common between the Brownian motion model and piece-wise linear regression.

\begin{table}[htbp]
	
	\begin{center}
		\begin{tabular}{|c|c|c|}
			\hline
			\cline{2-3} 
			\textbf{Brownian Motion} & \textbf{\textit{Predicted: 0}}& \textbf{\textit{Predicted: 1}}\\
			\hline
			Actual: 0& TN = 271& FP = 213 \\
			\hline
			Actual: 1& FN = 56 & FP = 219 \\
			\hline
		\end{tabular}
		\label{tab1}
		\caption{Brownian motion results}
	\end{center}
\end{table}	
	
\begin{table}[htbp]
	\begin{center}
		\begin{tabular}{|c|c|c|}
			\hline
			\cline{2-3} 
			\textbf{Brownian Motion} & \textbf{\textit{Predicted: 0}}& \textbf{\textit{Predicted: 1}}\\
			\hline
			Actual: 0& TN = 185& FP = 152 \\
			\hline
			Actual: 1& FN = 42 & FP = 147 \\
			\hline
		\end{tabular}
		\label{tab1}
		\caption{Brownian motion results - 526 systems}
	\end{center}
\end{table}	

\begin{table}[htbp]
	\begin{center}
		\begin{tabular}{|c|c|c|}
			\hline
			\cline{2-3} 
			\textbf{Brownian Motion} & \textbf{\textit{Predicted: 0}}& \textbf{\textit{Predicted: 1}}\\
			\hline
			Actual: 0& TN = 223& FP = 115 \\
			\hline
			Actual: 1& FN = 75 & FP = 113 \\
			\hline
		\end{tabular}
		\label{tab1}
		\caption{Piecewise Linear Regression - 526 systems}
	\end{center}
\end{table}

\item Considering systems that are common across both methods, the accuracy for Brownian motion model is 63\% and the accuracy for PLR is also 63\%. 
\end{itemize}

While the accuracy estimates for both systems are the same, the benefit of Brownian motion method comes from differences in true positives, false positives and the flexibility in business outcomes. The following sections expands on optimizing business decisions utilizing these statistical measures. 

\vspace{5mm}
\section{Optimizing Business Decisions}

Translating our model and results to business decisions, we considered the possible outcomes computed from the statistical measures. The business goal is to identify customers that are at risk of running out of capacity and proactively have customer support reach out  and discuss possible solutions. A more reactive approach would involve waiting until a full capacity event and a customer contacting support. This could involve more serious implications such as data loss on reaching full capacity. We can identify the business outcomes for a given system as:

\vspace{5mm}
$O(x) = b_{call}\ or\ b_{case}$

\vspace{5mm}
Each outcome has its associated dollar amount. While we are not at liberty to reveal our internal financial calculations, lets assume that the cost associated with calling a customer is $C(b_{call})$ and the cost associated with a customer filing a case is $C(b_{case})$. 

We identify the association between business outcome, $b_{call}$ or $b_{case}$ and the statistical measure computed. When the model suggests that the customer is at risk of running out of capacity i.e. $d(x) = 1$, this information can be passed on to customer support resulting in $O(x) = b_{call}$. From the statistical measures estimated, true positives (TP) and false positives (FP) include systems that are expected to be at risk. This can be represented as: 

\vspace{5mm}
$O(x) = b_{call} = TP + FP$

\vspace{5mm}
When the model does not identify a system that is at risk, to be at risk i.e. $d(x) = 0$, these are systems that are classified as true negatives (TN). If a system at risk is not identified as at risk, it results in the customer creating a case with a support engineer. This can be represented as: 

\vspace{5mm}
$O(x) = b_{case} = FN$

\vspace{5mm}

A model must be selected such that it minimizes the overall cost associated with both business outcomes i.e.

\vspace{5mm}
$Min C(O) = b_{call} *  C(b_{call}) + b_{case} *  C(b_{case})$

\vspace{5mm}
This is a convex optimization problem, where the minimum value of $C(O)$ provides an optimal business outcome. Figure 6 further explains the relationship. Zero phone calls are associated with a high cost to the business due to higher number of cases filed by customers. On the other end of the spectrum, more phone calls are also associated with increased costs. The minimum along this curve corresponds to an optimal business outcome that minimizes the overall cost.

\begin{figure}[htbp]
	\includegraphics[width=80mm,scale=0.8]{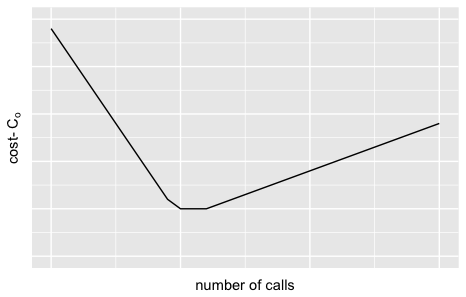}
	\caption{Convex optimization plot for Cost as a function of number of calls}
	\label{fig}
\end{figure}

When comparing costs associated with the piece-wise linear regression and the Brownian motion model, we observed a 39\% reduction in costs for Nutanix systems.

\vspace{5mm}
The Brownian motion model is beneficial in comparison to a piece-wise linear regression or ensemble method as:

\begin{itemize}
\item Discontinuous systems - The model can be applied to a larger set of systems, exhibiting a range of growth trends, in comparison to other methods. This is a result of variance and drift included as parameters within the model, providing better adaptability to a range of capacity patterns
\item Moves towards risk assessment - The limitation of applying a single forecasting method across systems that exhibit varied behavior, results in a poor forecast estimate and low credibility in the program. SER provides an opportunity to reframe the problem and think in terms of risk associated with running out of capacity
\item Hitting time problem - Instead of providing a point estimate, SER estimates the likelihood that a system would run out of capacity within a certain time period. Which implies that the system could reach full capacity at any point of time within the range we are computing for with a certain determined probability 
\item Personalized risk assessment -  SER  can be personalized based on a systems utilization and risk aversion preferences. A company with higher risk appetite might be able to withstand a 70\% risk of running out of capacity, versus a company with lower risk appetite where the costs associated with running out of capacity are much higher than purchasing more storage 
\item Minimizing cost - A comparison, between the Brownian motion model and regression, in terms of statistical measures such as true positive rate, false positive rate, provides an opportunity to optimize for a business decision of reducing costs associated. Based on preferences for true positives and false positives, the Brownian motion model can be optimized to reduce overall business costs across support and sales functions
\end{itemize}

\vspace{5mm}

\section{Conclusion and future work}

Our study showed how a naive forecasting method performs better than current capacity forecasting models, such as a peice-wise linear regression and forecasting ensembles. This provided motivation to reframe the problem and adopt a probabilistic approach to managing capacity. Stochastic Estimated Risk uses a Brownian motion with drift model to estimate the likelihood that systems would run out of capacity within a certain time frame.

In future work, we would like to examine the following areas to improve the model:

\begin{itemize}
	\item The model uses two parameters while estimating the likelihood, drift and variance. There is an opportunity to include more parameters that are reflective of storage environments
	\item The model has been applied solely to capacity utilization. We would like to examine additional workloads such as CPU 
\end{itemize}

 Comparing the Brownian motion method to existing forecasting models, we showed accuracy estimates for both samples were comparable. However, when translating our model to business decisions for Nutanix systems, implementing the Brownian motion model reduced costs by 39\% when compared to a piece-wise regression model. By reframing the existing problem statement and incorporating a wider range of systems, our approach ties the model with business outcomes and provides a cost efficient method to managing capacity. We hope this novel approach will be helpful to future researchers building models that focus on managing capacity utilization. 
 
 \vspace{5mm}

\section{} 
\vspace{12pt}

\end{document}